\documentclass[paper=a4, fontsize=11pt]{scrartcl}
\usepackage{graphicx}
\usepackage{amssymb}
\begin{document}
\begin{center}{\large{Topics in Non-Riemannian Geometry}}
\end{center}
\vspace*{1.5cm}
\begin{center}
A. C. V. V. de Siqueira
$^{*}$ \\
Retired \\
Universidade Federal Rural de Pernambuco \\
52.171-900, Recife, PE, Brazil.\\
${}^*$ E-mail:antonio.vsiqueira@ufrpe.br
\end{center}
\vspace*{1.5cm}
\begin{center}
{ Abstract}
\end{center}
\vspace*{1.0cm}
\begin{center}
In this paper, we present some new results on non-Riemannian geometry, more specifically, asymmetric connections and Weyl's geometry. For asymmetric connections, we show that a projective change in the symmetric part generates a vector field that is not arbitrary, as usually presented, but rather, the gradient of a non-arbitrary scalar function. We use normal coordinates for the symmetric part of asymmetric connections as well as for Weyl's geometry. This has a direct impact on asymmetric connections, although normal frames are usual in antisymmetric connections, unlike normal coordinates. In the symmetric part of asymmetric connections, the vector field obeys a well-known partial differential equation, whereas in Weyl's geometry, gauge vector fields obey an equation that we believe is presented for the first time in this paper. We deduce the exact solution of each of these vector fields as the gradient of a scalar function. For both asymmetric and Weyl's symmetric connections, the respective scalar functions obey respective scalar partial differential equations. As a consequence, Weyl's geometry is a conformal differential geometry and is associated with asymmetric geometry by a projective change. We also show that a metric tensor naturally appears in asymmetric geometry and is not introduced via a postulate, as is usually done. In Weyl's geometry, the electromagnetic gauge is the gradient of a non-arbitrary scalar function and electromagnetic fields are null. Despite the origin in Weyl's differential geometry, the use of the electromagnetic gauge is correct in Lagrangean and Hamiltonian formulations of field theories. 
\end{center}
\newpage
\section{Introduction}
In this paper, we use normal coordinates to show that a projective change in the symmetric part of asymmetric connections as well as Weyl's geometry for symmetric connections generate vector fields that are gradients of a non-arbitrary scalar function.
\newline 
Scalar functions are invariant due to coordinate transformations. Thus, our results in normal coordinates are also correct in other coordinates. 
\newline
Weyl's differential geometry is very important and originated the concept of gauge fields in electromagnetism and field theories. In Weyl's geometry, gauge field is the gradient of a scalar function and this implies
it will be a conformal differential geometry. It is less general than usually presented. 
\newline
This paper is organized as follows: In Sec. $2$, we define asymmetric connections and some geometric objects. We use a classical approach following L. P. Eisenhart,\cite{1}, because it is easy to read and has greater detail than offered in the present paper. In Sec. $3$, we present two different asymmetric connections that have the same paths and interrelated. In Sec. $4$, we use normal coordinates for the symmetric part of asymmetric connections and show that, for a projective change in connections, the vector field is the gradient of a non-arbitrary scalar function. In Sec. $5$, we present some results of Weyl's differential geometry. Using  normal coordinates, we show that the most general gauge field is the gradient of a non-arbitrary scalar function, showing that Weyl's geometry is a conformal geometry. In Sec. $6$, we show that a projective change in the symmetric part of asymmetric connections is equivalent to a conformal transformation in the metric tensor of Weyl's geometry. The influence of Weyl's gauge field on the change in the symmetric and antisymmetric parts of asymmetric connections is explicitly shown. In Sec. $7$, we present our concluding remarks.
\newpage
\renewcommand{\theequation}{\thesection.\arabic{equation}}
\section{Asymmetric Connections}
\setcounter{equation}{0}
 $         $
In this section, we introduce asymmetric connections and some associated geometric objects.
\newline
An asymmetric connection can be defined as the sum of a symmetric connection with another antisymmetric connection, as follows
\begin{equation}
L_{\mu \nu}^{\eta}=\Gamma_{\mu \nu}^{\eta}+\Omega_{\mu \nu}^{\eta},
\end{equation} 
\begin{equation}
 \Gamma_{\mu \nu}^{\eta} =\Gamma_{\nu \mu}^{\eta},
\end{equation}
and
\begin{equation}
\Omega_{\nu \mu}^{\eta}=-\Omega_{\mu \nu}^{\eta}.
\end{equation}
We will now present the covariant derivatives for covariant components of a vector for each different connection
\begin{equation}
X_{\mu \shortmid\nu}= \partial_{\nu} X_{\mu
} -L_{\mu \nu}^{\eta}X_{\eta},
\end{equation} 
\begin{equation}
X_{\mu ;\nu}= \partial_{\nu} X_{\mu
} -\Gamma_{\mu \nu}^{\eta}X_{\eta},
\end{equation} 
\begin{equation}
X_{\mu ,\nu}= \partial_{\nu} X_{\mu
} -C_{\mu \nu}^{\eta}X_{\eta},
\end{equation} 
where $C_{\mu \nu}^{\eta}$ are the components of the Christoffel symbols.
\newline
Let us present the following curvatures for the different connections
\begin{equation}
L^{\alpha}_{\mu \sigma \nu }=-\partial_{\nu} L_{\mu \sigma
}^{\alpha}+\partial_{\sigma} L_{\mu \nu}^{\alpha}-L_{\mu
\sigma}^{\eta} L_{n \nu}^{\alpha}+L_{\mu \nu}^{\eta} L
_{\sigma \eta}^{\alpha},
\end{equation}
\begin{equation}
B^{\alpha}{}_{\mu \sigma \nu }=-\partial_{\nu} \Gamma_{\mu \sigma
}^{\alpha}+\partial_{\sigma}\Gamma_{\mu \nu}^{\alpha}-\Gamma_{\mu
\sigma}^{\eta}\Gamma_{n \nu}^{\alpha}+\Gamma_{\mu \nu}^{\eta}\Gamma
_{\sigma \eta}^{\alpha},
\end{equation}
\begin{equation}
\Omega^{\alpha}{}_{\mu \sigma \nu }=-\Omega_{\mu \sigma\shortmid\nu
}^{\alpha} +\Omega_{\mu \nu\shortmid\sigma }^{\alpha}+\Omega_{\mu
\sigma}^{\eta} \Omega_{n \nu}^{\alpha}-\Omega_{\mu \nu}^{\eta} \Omega
_{\sigma \eta}^{\alpha},
\end{equation}
\begin{equation}
R^{\alpha}{}_{\mu \sigma \nu }=-\partial_{\nu} C_{\mu \sigma
}^{\alpha}+\partial_{\sigma} C_{\mu \nu}^{\alpha}-C_{\mu
\sigma}^{\eta} C_{n \nu}^{\alpha}+C_{\mu \nu}^{\eta} C
_{\sigma \eta}^{\alpha},
\end{equation}
with Ricci's tensor
\begin{equation}
R_{\mu \nu}=R^{\alpha}{_{\mu \nu \alpha }}.
\end{equation}
The parallel transport is given by
\begin{equation}
 \frac{dX^{\eta}}{dt} + L_{\mu \nu}^{\eta}X^{\mu} \frac{dx^{\nu}}{dt}=0.
\end{equation}
In the next section, we present a more general concept of parallelism.
\newpage
\renewcommand{\theequation}{\thesection.\arabic{equation}}
\section{Changes in Connections that Preserve Parallelism }
\setcounter{equation}{0}
 $         $
Here, we present a more general concept of parallelism between two vectors and how the same path is possible for two different asymmetric connections.
\newline
Let us consider two vector fields with the same direction at each point where the respective components satisfy the following relationship
\begin{equation}
\widehat{X}^{\eta}=\varphi X^{\eta}.
\end{equation}
Each component is parallel transported by a different asymmetrical connection as follows
\begin{equation}
 \frac{d\widehat{X}^{\eta}}{dt} + \widehat{L}_{\mu \nu}^{\eta}\widehat{X}^{\mu} \frac{dx^{\nu}}{dt}=0.
\end{equation}
\begin{equation}
 \frac{dX^{\eta}}{dt} + L_{\mu \nu}^{\eta}X^{\mu} \frac{dx^{\nu}}{dt}=0.
\end{equation}
 Deriving (3.1) and using
\begin{equation}
 L_{\mu \nu}^{\eta}\widehat{X}^{\mu} \frac{dx^{\nu}}{dt}=\varphi L_{\mu \nu}^{\eta}X^{\mu} \frac{dx^{\nu}}{dt},
\end{equation}
We have
\begin{equation}
 \frac{d\widehat{X}^{\eta}}{dt} + {L}_{\mu \nu}^{\eta}\widehat{X}^{\mu} \frac{dx^{\nu}}{dt}=f(t)\widehat{X}^{\eta},
\end{equation}
where (3.3) was used and
\begin{equation}
 f(t)=\frac{1}{\varphi}\frac{d\varphi}{dt}.
\end{equation}
\newline
Let us consider the following multiplication
\begin{equation}
\widehat{X}^{\sigma} \frac{d\widehat{X}^{\eta}}{dt} + {L}_{\mu \nu}^{\eta}\widehat{X}^{\sigma}\widehat{X}^{\mu} \frac{dx^{\nu}}{dt}=f(t)\widehat{X}^{\sigma}\widehat{X}^{\eta},
\end{equation}
\begin{equation}
\widehat{X}^{\eta}\frac{d\widehat{X}^{\sigma}}{dt} + {L}_{\mu \nu}^{\sigma}\widehat{X}^{\eta}\widehat{X}^{\mu}\frac{dx^{\nu}}{dt}=f(t)\widehat{X}^{\eta}\widehat{X}^{\sigma}.
\end{equation}
We then have
\begin{equation}
\widehat{X}^{\sigma} \frac{d\widehat{X}^{\eta}}{dt} + {\widehat{L}}_{\mu \nu}^{\eta}\widehat{X}^{\sigma}\widehat{X}^{\mu} \frac{dx^{\nu}}{dt}=0,
\end{equation}
\begin{equation}
\widehat{X}^{\eta}\frac{d\widehat{X}^{\sigma}}{dt} + {\widehat{L}}_{\mu \nu}^{\sigma}\widehat{X}^{\eta}\widehat{X}^{\mu} \frac{dx^{\nu}}{dt}=0.
\end{equation}
Eliminating $f(t)$ in (3.7) and (3.8), from (3.9) and (3.10), we have
\begin{equation}
\widehat{X}^{\eta}\lbrace\frac{d\widehat{X}^{\sigma}}{dt} + {L}_{\mu \nu}^{\sigma}\widehat{X}^{\mu}\frac{dx^{\nu}}{dt}\rbrace-\widehat{X}^{\sigma}\lbrace\frac{d\widehat{X}^{\eta}}{dt} + {L}_{\mu \nu}^{\eta}\widehat{X}^{\mu}\frac{dx^{\nu}}{dt}\rbrace=0,
\end{equation}
\begin{equation}
\widehat{X}^{\eta}\lbrace\frac{d\widehat{X}^{\sigma}}{dt} + {\widehat{L}}_{\mu \nu}^{\sigma}\widehat{X}^{\mu}\frac{dx^{\nu}}{dt}\rbrace-\widehat{X}^{\sigma}\lbrace\frac{d\widehat{X}^{\eta}}{dt} + {\widehat{L}}_{\mu \nu}^{\eta}\widehat{X}^{\mu}\frac{dx^{\nu}}{dt}\rbrace=0.
\end{equation}
From (3.11) and (3.12), we get
\begin{equation}
({{{\widehat{L}}_{\mu \nu}^{\sigma}-{L}_{\mu \nu}^{\sigma}}})\widehat{X}^{\eta}\widehat{X}^{\mu}\frac{dx^{\nu}}{dt}=({{\widehat{L}}_{\mu \nu}^{\eta}-{L}_{\mu \nu}^{\eta}})\widehat{X}^{\sigma}\widehat{X}^{\mu}\frac{dx^{\nu}}{dt}.
\end{equation}
We now define the following relation between the connections
\begin{equation}
{\widehat{L}}_{\mu \nu}^{\sigma}-{L}_{\mu \nu}^{\sigma}={a}_{\mu \nu}^{\sigma}.
\end{equation}
After some simple calculations, we have
\begin{equation}
 n{a}_{\mu \nu}^{\sigma}={\delta}_{\mu}^{\sigma}{a}_{\lambda\mu}^{\lambda}.
\end{equation}
We now define
\begin{equation}
 2n{\psi}_{\mu}={a}_{\lambda\mu}^{\lambda}.
\end{equation}
Using (3.15) and (3.16) in (3.14), we get
\begin{equation}
{\widehat{L}}_{\mu \nu}^{\sigma}={L}_{\mu \nu}^{\sigma}+2{\delta}_{\mu}^{\sigma}{\psi}_{\nu}.
\end{equation}
Using (3.18)in(3.5)
\begin{equation}
\widehat{\Gamma}_{\mu \nu}^{\eta}=\Gamma_{\mu \nu}^{\eta}+{\delta}_{\mu}^{\eta}{\psi}_{\nu}+{\delta}_{\nu}^{\eta}{\psi}_{\mu},
\end{equation} 
and
\begin{equation}
\widehat{\Omega}_{\mu \nu}^{\eta}=\Omega_{\mu \nu}^{\eta}+{\delta}_{\mu}^{\eta}{\psi}_{\nu}-{\delta}_{\nu}^{\eta}{\psi}_{\mu}.
\end{equation} 
Using (3.17) and (2.7), we get
\begin{equation}
\widehat{L}^{\alpha}_{\mu \sigma \nu }=L^{\alpha}_{\mu \sigma \nu }-2{\delta}_{\mu}^{\alpha}(\frac{\partial\psi_{\sigma}}{\partial{x^{\nu}}}-\frac{\partial\psi_{\nu}}{\partial{x^{\sigma}}}),
\end{equation}
Different connections that change according to (3.17), (3.18) and (3.19) have the same path and generate curvatures that interrelate through (3.20).
\renewcommand{\theequation}{\thesection.\arabic{equation}}
\section{Normal Coordinates and Projective Change of Connections}
\setcounter{equation}{0}
In this section, we consider normal coordinates and show that, for a projective change in connections, the vector field is actually the gradient of a scalar function.
\newline
Let us consider the symmetric part of two asymmetric connections given by (3.18) as functions of the variables $(x^{\eta})$.
\newline
For two normal coordinates, equations of paths through the origin are given by \cite{1},
\begin{equation}
 y^{\eta}=\frac{dx^{\eta}}{dt}t, 
\end{equation}
\begin{equation}
 z^{\eta}=\frac{dx^{\eta}}{ds}s, 
\end{equation}
with evolution parameters t and s.
\newline
There is a simple relationship between the normal coordinates given by
\begin{equation}
 y^{\eta}=\frac{z^{\eta}}{f(z)}, 
\end{equation}
\begin{equation}
 z^{\eta}=\frac{y^{\eta}}{g(y)}, 
\end{equation}
subject to the condition
\begin{equation}
 f(z)g(y)=1. 
\end{equation}
For normal coordinates $ y^{\eta}$ and $z^{\eta}$, (3.18) will be given by
\begin{equation}
\widehat{\Sigma}_{\mu \nu}^{\eta}=\Sigma_{\mu \nu}^{\eta}+{\delta}_{\mu}^{\eta}{\psi}_{\nu}+{\delta}_{\nu}^{\eta}{\psi}_{\mu},
\end{equation}
and obeys the following conditions
\begin{equation}
\widehat{\Sigma}_{\mu \nu}^{\eta} y^{\mu}y^{\nu}=0,
\end{equation}
\begin{equation}
\Sigma_{\mu \nu}^{\eta}z^{\mu}z^{\nu}=0.
\end{equation}
After some considerations, we have ( \cite{1}, Sec. $33$, eq. (33.6))
\begin{equation}
(z^{\mu}\psi_{\mu}+f^{-1}\frac{\partial{f}}{\partial{z^{\nu}}}z^{\nu})(f-\frac{\partial{f}}{\partial{z^{\sigma}}}z^{\sigma})+\frac{1}{2}f_{,\mu\nu}z^{\mu}z^{\nu}=0,
\end{equation}
$f_{,\mu\nu}$ is the second partial derivative of f.
\newline
The exact solution of (4.9) is given by
\begin{equation}
2\psi_{\mu}=\partial_{\mu}log(f^{-1}+\frac{\partial{f^{-1}}}{\partial{z^{\sigma}}}z^{\sigma}).
\end{equation}
This implies that $\psi_{\mu}$ is derived from a scalar $\psi$ given by
\begin{equation}
2\psi=log(f^{-1}+\frac{\partial{f^{-1}}}{\partial{z^{\sigma}}}z^{\sigma}).
\end{equation}
We can put
\begin{equation}
\psi_{\mu}=\partial_{\mu}\psi.
\end{equation}
Let us now define
\begin{equation}
f^{-1}=h,
\end{equation}
and substituting (4.13) in (4.11), we have
\begin{equation}
h+\frac{\partial{h}}{\partial{z^{\sigma}}}z^{\sigma}=e^{2\psi}.
\end{equation}
When $\psi$ is a function only of $\frac{\partial{h}}{\partial{z^{\sigma}}}$, (4.14) is the generalized Clairaut's equation, \cite{2}. 
Substituting (4.12) into (3.17), (3.18), (3.19) and (3.20), we have
\begin{equation}
{\widehat{L}}_{\mu \nu}^{\sigma}={L}_{\mu \nu}^{\sigma}+2{\delta}_{\mu}^{\sigma}\partial_{\mu}\psi,
\end{equation}
\begin{equation}
\widehat{\Gamma}_{\mu \nu}^{\eta}=\Gamma_{\mu \nu}^{\eta}+{\delta}_{\mu}^{\eta}\partial_{\nu}\psi+{\delta}_{\nu}^{\eta}\partial_{\mu}\psi,
\end{equation} 
\begin{equation}
\widehat{\Omega}_{\mu \nu}^{\eta}=\Omega_{\mu \nu}^{\eta}+{\delta}_{\mu}^{\eta}\partial_{\nu}\psi-{\delta}_{\nu}^{\eta}\partial_{\mu}\psi,
\end{equation} 
and
\begin{equation}
\widehat{L}^{\alpha}_{\mu \sigma \nu }=L^{\alpha}_{\mu \sigma \nu },
\end{equation}
which implies a simplification of (3.20). 
\newline
Other geometric objects not present in this section also change in the two geometries.
\renewcommand{\theequation}{\thesection.\arabic{equation}}
\section{Weyl's Differential Geometry}
\setcounter{equation}{0}
Weyl's differential geometry is very important and originated the concept of gauge fields in electromagnetism. Weyl's geometry reduced to a conformal differential geometry when a gauge field is the gradient of a scalar function \cite{3}. In this section, we show that the most general gauge field is the gradient of a scalar function that obeys a first-order partial differential equation.
\newline
In Weyl's differential geometry, the derivative of the metric tensor obeys the following expression
\begin{equation}
 g_{\mu \nu};{\eta}+2g_{\mu \nu}{\Phi}_{\eta}=0,
\end{equation} 
and the connection is given by
\begin{equation}
{W}_{\mu \nu}^{\eta}=C_{\mu \nu}^{\eta}+{\delta}_{\mu}^{\eta}{\Phi}_{\nu}+{\delta}_{\nu}^{\eta}{\Phi}_{\mu}-g_{\mu \nu}{\Phi}^{\eta},
\end{equation}
with $C_{\mu \nu}^{\eta}$ as components of Christoffel's symbols.
\newline
With a similar treatment used in obtaining (4.9) and $ g_{\mu \nu}{z}^{\mu}{z}^{\nu}\neq0 $ (a not light-like interval), we get 
\begin{eqnarray}
2z^{\mu}\Phi_{\mu}z^{\alpha}(f^{-1}-f^{-2}\frac{\partial{f}}{\partial{z^{\nu}}}z^{\nu})-z_{\sigma}z^{\sigma}(\Phi^{\alpha}f^{-1}\\
 \nonumber
-f^{-2}z^{\alpha}\frac{\partial{f}}{\partial{z^{\nu}}}\Phi^{\nu})+z^{\alpha}f^{-2}(2\frac{\partial{f}}{\partial{z^{\nu}}}z^{\nu}+f_{,\mu\nu}z^{\mu}z^{\nu}-2f^{-1}\frac{\partial{f}}{\partial{z^{\nu}}}z^{\nu}\frac{\partial{f}}{\partial{z^{\mu}}}z^{\mu})=0.
\end{eqnarray}
The exact solution of (5.1) is given by
\begin{equation}
\Phi_{\mu}=-2\partial_{\mu}logf+\partial_{\mu}log(f-\frac{\partial{f}}{\partial{z^{\sigma}}}z^{\sigma}).
\end{equation}
This implies that $\Phi_{\mu}$ is derived from a scalar $\Phi$ given by
\begin{equation}
\Phi=-2logf+log(f-\frac{\partial{f}}{\partial{z^{\sigma}}}z^{\sigma})+ const.
\end{equation}
Substituting (4.13) in (5.5), we get
\begin{equation}
h+\frac{\partial{h}}{\partial{z^{\sigma}}}z^{\sigma}=e^{\Phi}.
\end{equation}
The verification that (4.10) is the solution of (4.9) is simple, but that (5.4) is the solution of (5.3) is more complex. For this, we need to put (5.3) in the form
\begin{equation}
\frac{\partial{f}}{\partial{z^{\sigma}}}(.........)=0,
\end{equation}
after the use of scalar product $z_{\sigma}z^{\sigma}\neq0$.
\newline
Comparing (4.14) with (5.6), we see that $2\psi+const.=\Phi$
\newline
or
\begin{equation}
2\frac{\partial{\psi}}{\partial{z^{\sigma}}}=\frac{\partial{\Phi}}{\partial{z^{\sigma}}}.
\end{equation}
Let us consider a light-like interval, $z_{\sigma}z^{\sigma}=0$, for Weyl's gauge. 
\newline
For this type of interval in place of $\frac{\partial{\Phi}}{\partial{z^{\sigma}}}$, put $\frac{\partial{\phi}}{\partial{z^{\sigma}}}$. We then get
\begin{equation}
(z^{\mu}\phi_{\mu}+f^{-1}\frac{\partial{f}}{\partial{z^{\nu}}}z^{\nu})(f-\frac{\partial{f}}{\partial{z^{\sigma}}}z^{\sigma})+\frac{1}{2}f_{,\mu\nu}z^{\mu}z^{\nu}=0.
\end{equation}
The connection is given by (5.2) and the development of the equation for $\phi$ for a light-like interval leads directly to (5.9), which is identical to (4.9). If we assume a light-like interval in (5.3), we immediately get (5.9), but the correct procedure would be to assume the light-like interval from the beginning.
\newline
As with (4.9), the exact solution of (5.9) is given by
\begin{equation}
2\phi_{\mu}=\partial_{\mu}log(f^{-1}+\frac{\partial{f^{-1}}}{\partial{z^{\sigma}}}z^{\sigma}).
\end{equation}
This implies 
\begin{equation}
2\phi=log(f^{-1}+\frac{\partial{f^{-1}}}{\partial{z^{\sigma}}}z^{\sigma}).
\end{equation}
We can state
\begin{equation}
\phi_{\mu}=\partial_{\mu}\phi.
\end{equation}
In (5.11), using
\begin{equation}
f^{-1}=h,
\end{equation}
we get
\begin{equation}
h+\frac{\partial{h}}{\partial{z^{\sigma}}}z^{\sigma}=e^{2\phi}.
\end{equation}
Comparing (4.14) with (5.14), we see that $\psi+const.=\phi$
\newline
or
\begin{equation}
\frac{\partial{\psi}}{\partial{z^{\sigma}}}=\frac{\partial{\phi}}{\partial{z^{\sigma}}}.
\end{equation}
We conclude for $z_{\sigma}z^{\sigma}\neq0$, the gauge field of Weyl's geometry is twice the vector field of the projective change and, for a light-like interval, $z_{\sigma}z^{\sigma}=0$, vector fields are equal.
\newline
These scalar fields are invariant under coordinate transformations and therefore apply to another coordinate system related to normal coordinates. We have two differential geometries. Weyl's differential geometry is a conformal geometry, while the another conserves paths for different connections.
\newpage
\renewcommand{\theequation}{\thesection.\arabic{equation}}
\section{Association Between Asymmetric and Weyl's Connections}
\setcounter{equation}{0}
In the previous sections, we showed that vector fields obtained by a projective change in the symmetrical part of asymmetric connections and the gauge fields in Weyl's theory are related by the equation (5.8) if the interval is not light-like, $z_{\sigma}z^{\sigma}\neq0$, and are related by (5.15) for a light-like interval, $z_{\sigma}z^{\sigma}=0$.
\newline
Let us consider a interval that is not light-like. From (4.16), (5.2) and (5.8), we get
\begin{equation}
2\widehat{\Gamma}_{\mu \nu}^{\eta}=2\Gamma_{\mu \nu}^{\eta}+ +{\delta}_{\mu}^{\eta}\partial_{\nu}\Phi+{\delta}_{\nu}^{\eta}\partial_{\mu}\Phi,
\end{equation}
which is the symmetric part of the projective connection as a function of Weyl's gauge field.
\begin{equation}
\widehat{\Gamma}_{\mu \nu}^{\eta}=\Gamma_{\mu \nu}^{\eta}+2{W}_{\mu \nu}^{\eta}-2C_{\mu \nu}^{\eta}+\frac{1}{2}g_{\mu \nu}{\Phi}^{\eta},
\end{equation}
which is the symmetric part of the projective connection as a function of Weyl's connection and Weyl's gauge field.
\begin{equation}
2\widehat{\Omega}_{\mu \nu}^{\eta}=2\Omega_{\mu \nu}^{\eta}+{\delta}_{\mu}^{\eta}\partial_{\nu}\Phi-{\delta}_{\nu}^{\eta}\partial_{\mu}\Phi,
\end{equation} 
which is the antisymmetric part of an asymmetric connection as a function of Weyl's gauge field.
\newline
Equations (6.1) and (6.2) show that a projective change in the symmetric part of asymmetric connections is equivalent to a conformal transformation in the metric tensor. Equation (6.3) explicitly shows the influence of Weyl's gauge field on the change in the antisymmetric part of asymmetric connections. From (6.2), it is a natural process to introduce a metric in asymmetric geometry.
\newline
It is evident that (4.12) implies a simplification in Weyl's projective tensor, although Weyl's conformal tensor does not change. These tensors are defined in \cite{1} and \cite{3}). 
\newline
Let us write Weyl's connection and rewrite (6.1), (6.3) and (5.6)
\begin{equation}
{W}_{\mu \nu}^{\eta}=C_{\mu \nu}^{\eta}+{\delta}_{\mu}^{\eta}{\Phi}_{\nu}+{\delta}_{\nu}^{\eta}{\Phi}_{\mu}-g_{\mu \nu}{\Phi}^{\eta},
\end{equation}
\begin{equation}
h+\frac{\partial{h}}{\partial{z^{\sigma}}}z^{\sigma}=e^{\Phi},
\end{equation}
\begin{equation}
2\widehat{\Gamma}_{\mu \nu}^{\eta}=2\Gamma_{\mu \nu}^{\eta}+ +{\delta}_{\mu}^{\eta}\partial_{\nu}\Phi+{\delta}_{\nu}^{\eta}\partial_{\mu}\Phi,
\end{equation}
\begin{equation}
2\widehat{\Omega}_{\mu \nu}^{\eta}=2\Omega_{\mu \nu}^{\eta}+{\delta}_{\mu}^{\eta}\partial_{\nu}\Phi-{\delta}_{\nu}^{\eta}\partial_{\mu}\Phi.
\end{equation} 
Let us consider a light-like interval. The relation between the fields of the two geometries is given by (5.15) and we get
\begin{equation}
{W}_{\mu \nu}^{\eta}(L)=C_{\mu \nu}^{\eta}+{\delta}_{\mu}^{\eta}{\phi}_{\nu}+{\delta}_{\nu}^{\eta}{\phi}_{\mu}-g_{\mu \nu}{\phi}^{\eta},
\end{equation}
\begin{equation}
h+\frac{\partial{h}}{\partial{z^{\sigma}}}z^{\sigma}=e^{2\phi},
\end{equation}
\begin{equation}
\widehat{\Gamma}_{\mu \nu}^{\eta}=\Gamma_{\mu \nu}^{\eta}+ +{\delta}_{\mu}^{\eta}\partial_{\nu}\phi+{\delta}_{\nu}^{\eta}\partial_{\mu}\phi,
\end{equation}
\begin{equation}
\widehat{\Omega}_{\mu \nu}^{\eta}=\Omega_{\mu \nu}^{\eta}+{\delta}_{\mu}^{\eta}\partial_{\nu}\phi-{\delta}_{\nu}^{\eta}\partial_{\mu}\phi,
\end{equation}
in which ${W}_{\mu \nu}^{\eta}(L)$ is Weyl's connection for a light-like interval.
\newline
Connections (6.4) and (6.8) have different geodesics and curvatures \cite{3}.
\newline
This produces different effects on Weyl's conformal tensor, projective tensor and asymmetric connections.
\newline
For an interval that is not light-like, we can use equation (6.2) to put Weyl's conformal tensor as a function of Weyl's projective tensor and, for a light-like interval, we can use an equation equivalent to (6.2).
\newline
Weyl's conformal tensor is a function of pseudo-Riemannian geometric objects, specifically Riemann's tensor, Ricci's tensor and curvature. We can then put Weyl's projective tensor as a function of a pseudo-Riemannian geometry.
\newline
In a pseudo-Riemannian geometry, we can define a Fermi transport for an interval that is not light-like or is light-like on a curve. This can be extended to other symmetric connections, such as Weyl's, and it is possible to choose coordinates in which the connections are zero at all points of a curve or a portion of it \cite{1}. 
\section{Concluding Remarks}
\setcounter{equation}{0}
In this paper, we have used normal coordinates to show that a projective change in the symmetric part of asymmetric connections as well as Weyl's geometry for symmetric connections generate vector fields that are gradients of non-arbitrary scalar functions. Consequently, Weyl's geometry is a conformal differential geometry and is associated with an asymmetric geometry through a projective change in the symmetric part. We also show that a metric tensor naturally appears in asymmetric geometry and is not introduced via a postulate, as is usually done. In other words, if we want to introduce a metric into an asymmetric geometry without postulating it, the natural way is to make a projective change in the symmetric part of the asymmetric connection. This is equivalent to a conformal transformation of Weyl's metric.
\newline
Connections (6.4) and (6.8) respectively together with (6.5) and (6.9) have different geodesics and curvatures. For each value of the coordinates, this produces different effects in Weyl's projective tensor, Weyl's conformal tensor and asymmetric connections.
\newline
We believe that applications in physical theories will impose constraints on h or $\Phi$ in (6.5) and h or $\phi$ in (6.9). The first option would be to choose the function h as a constant plus a periodic function multiplied by a very small constant, with undetermined $\Phi$ and $\phi$. The second option would be to choose each $\Phi$ and $\phi$ as different functions given by a constant plus a periodic function multiplied by a very small constant, with h as an undetermined function. It also seems possible to choose h, $\Phi$ and $\phi$ as different functions, giving by a constant plus a periodic function multiplied by a very small constant. These assumptions will be discussed in an upcoming paper.
 
\end{document}